\documentclass[12pt,a4paper]{article}
\pdfoutput=1
\usepackage{bm, amssymb,pifont,cancel, amsmath,comment,color}
\usepackage{here}
\usepackage{cite}
\usepackage{graphicx}
\usepackage{subfigure}

\makeatletter

\@addtoreset{equation}{section}
\makeatother
 
\begin{document}

\begin{titlepage}
\null
\begin{flushright}
WU-HEP-19-06\\
KEK-TH-2137
\end{flushright}

\vskip 1cm
\begin{center}
\baselineskip 0.8cm
{\LARGE \bf
Behaviors of two supersymmetry breaking scales in $\mathcal{N}=2$ supergravity}

\lineskip .75em
\vskip 1cm

\normalsize

{\large } {\large Hiroyuki Abe} $^{1}${\def\thefootnote{\fnsymbol{footnote}}\footnote[1]{E-mail address: abe@waseda.jp}},
{\large Shuntaro Aoki} $^{1}${\def\thefootnote{\fnsymbol{footnote}}\footnote[2]{E-mail address: shun-soccer@akane.waseda.jp}}, 
{\large } {\large Sosuke Imai} $^{1}${\def\thefootnote{\fnsymbol{footnote}}\footnote[3]{E-mail address: s.i.sosuke@akane.waseda.jp}},
{\large } {\large Yutaka Sakamura} $^{2,3}${\def\thefootnote{\fnsymbol{footnote}}\footnote[4]{E-mail address: sakamura@post.kek.jp}}

\vskip 1.0em

$^1${\small\it Department of Physics, Waseda University, \\ 
Tokyo 169-8555, Japan}
\vskip 1.0em

$^2${\small{\it KEK Theory Center, Institute of Particle and Nuclear Studies, KEK,\\
1-1 Oho, Tsukuba, Ibaraki 305-0801, Japan}}
\vskip 1.0em

$^3${\small{\it Department of Particles and Nuclear Physics,\\
SOKENDAI (The Graduate University for Advanced Studies),\\
1-1 Oho, Tsukuba, Ibaraki 305-0801, Japan}}
\vskip 1.0em

\vspace{12mm}

{\bf Abstract}\\[5mm]
{\parbox{13cm}{\hspace{5mm} \small
We study the supersymmetry breaking patterns in four-dimensional $\mathcal{N}=2$ gauged supergravity.  The model contains multiple (Abelian) vector multiplets and a single hypermultiplet which parametrizes SO$(4,1)/{\rm{SO}}(4)$ coset. We derive the expressions of two gravitino masses under {\it{general}} gaugings and prepotential based on the embedding tensor formalism, and discuss their behaviors in some concrete models. Then we confirm that in a single vector multiplet case, the partial breaking always occurs when the third derivative of the prepotential exists at the vacuum, 
which is consistent with the result of Ref.~\cite{Antoniadis:2018blk}, but we can have several breaking patterns otherwise. The discussion is also generalized to the case of multiple vector multiplets, and we found that the full ($\mathcal{N}=0$) breaking occurs even if the third derivative of the prepotential is nontrivial.


}}

\end{center}

\end{titlepage}

\tableofcontents
\vspace{35pt}
\hrule

\section{Introduction}\label{intro}
Extended ($\mathcal{N}\geq 2$) supergravity in four dimensions naturally appears from higher-dimensional supergravity and string compactifications (see~\cite{Grana:2005jc,Blumenhagen:2006ci} for review). For phenomenological applications, we need to consider its breaking mechanism, since in extended supergravity there is no chiral-structure which is necessary to describe real world. As regards for the breaking of extended supergravity, there are several breaking patterns to be considered in contrast to $\mathcal{N}=1$ case. For example, the vacuum may preserve some supersymmetries partially. Also, even if the full breaking occurs at the vacuum, some of supersymmetry breaking scales may be degenerate or hierarchical. The purpose of this paper is to clarify the relations between the breaking patterns and input parameters in the theory, which is motivated mainly as follows:
\begin{itemize}
  \item 
  Indeed, some of these input parameters (e.g., gauge couplings) are determined by flux in the context of string compactifications. Therefore, it is necessary for understanding phenomenological/cosmological aspects of flux compactifications. 
  \item 
  The cases of the partial breaking where some supersymmetries remain unbroken, are studied well in both of the local~\cite{Ferrara:1995gu,Ferrara:1995xi,Fre:1996js,Itoyama:2006ef,Maruyoshi:2006te,Louis:2009xd,Louis:2010ui,Hansen:2013dda,Antoniadis:2018blk} and the global~\cite{Hughes:1986dn,Antoniadis:1995vb,Fujiwara:2005hj,Fujiwara:2005kf,Fujiwara:2006qh} cases. Their relations are discussed in Refs.~\cite{Ferrara:1995xi,David:2003dh,Andrianopoli:2015rpa,Laamara:2017hdl}. Those models evade the no-go theorem~\cite{Witten:1981nf,Cecotti:1984rk,Cecotti:1984wn}, and it is known that stable minima are ensured in this case~\cite{Louis:2009xd}.
  It is also discussed about some roles of the partial breaking in the effective description of D-branes (in $\mathcal{N}=2$ case, see~\cite{Bagger:1996wp,Rocek:1997hi,GonzalezRey:1998kh,Tseytlin:1999dj,Burgess:2003hx,Antoniadis:2008uk,Kuzenko:2009ym,Ambrosetti:2009za,Kuzenko:2011ya,Ferrara:2014oka,Kuzenko:2015rfx,Ferrara:2016crd,Kuzenko:2017zla,Antoniadis:2017jsk,Farakos:2018aml,Cribiori:2018jjh,Antoniadis:2019gbd} for example). Therefore, it is important to ask under what situations the partial breaking occurs. 
  \item 
  From more bottom-up perspectives, if there exists an extended supersymmetry and its breaking, additional massive modes we do not have in the usual $\mathcal{N}=1$ supergravity models necessarily appear and they could affect the cosmological history. For example, in $\mathcal{N}=2$ supergravity, we have double massive gravitinos when $\mathcal{N}=2\rightarrow 0$ breaking occurs. Their effects cannot be negligible if the two breaking scales are close to each other, and the usual $\mathcal{N}=1$ description might be broken down in that case. Then, it is interesting to investigate what difference and phenomenological consequence appear if such extended supersymmetry exists. To this end, we need to know precisely the breaking patterns, the resultant spectra, and coupling constants.
  
\end{itemize}

As a first step in this paper, we achieve our purpose by taking $\mathcal{N}=2$ supergravity in four dimensions as the simplest and concrete example. In particular, we focus on a model which contains multiple (Abelian) vector multiplets and a single hypermultiplet which parametrizes SO$(4,1)/{\rm{SO}}(4)$ coset. The isometries in the hyper sector are gauged by the vector fields in the vector multiplets as well as the graviphoton.

This model can be regarded as a multiple generalization of the vector sector of Refs.~\cite{Ferrara:1995gu,Ferrara:1995xi}, where it is shown that the $\mathcal{N}=0,1$ and $2$ vacua can be realized within the single framework, depending on the gauge couplings. Therefore, the model would be appropriate for considering various breaking patterns.\footnote{Besides this model, the full supersymmetry breaking models ($\mathcal{N}=2\rightarrow 0$) are discussed in  Refs.~\cite{Dudas:2017sbi,Kuzenko:2017gsc}, based on  $\mathcal{N}=2$ supergravity constrained superfield. Also, in Refs.~\cite{Antoniadis:2012cg,Laamara:2015rbq}, a model where $\mathcal{N}=2$ 
global supersymmetry can be broken at two different scales is discussed.}

Based on the setup above and under a specific gauging, we have explicitly constructed a model which interpolates $\mathcal{N}=0$ and
$\mathcal{N}=1$ Minkowski vacua, and evaluated the mass spectrum in our previous paper~\cite{Abe:2019svc}. Here we consider general gaugings extending our previous analysis. We employ the so-called embedding tensor formalism~\cite{deWit:2002vt,deWit:2005ub}, which allows us to treat the general gauging without changing duality frame (see~\cite{Samtleben:2008pe,Trigiante:2016mnt} for review). Then, we derive the general expressions of the two gravitino masses and study their behaviors by case analysis. As we will see, the breaking patterns are governed by the gauge couplings and the form of the prepotential. In particular, 
our approach reproduces the result of Ref.~\cite{Antoniadis:2018blk} in a systematic way, which claims, the partial breaking always occurs when the third derivative of the prepotential exists at the vacuum. Moreover, we explicitly show that there are several breaking patterns otherwise. We also discuss the case of multiple vector multiplets and investigate how the situation becomes different from the single case.  


The paper is organized as follows. In Sec.~\ref{setup}, we specify our model and introduce the notation used in the paper. Then, we evaluate the gravitino masses under the general gauging in Sec.~\ref{GM}. There, we briefly explain their behaviors and discuss the conditions to realize special cases such as $\mathcal{N}=1,2$ vacua. In Sec.~\ref{SCA}, we analyze the scalar potential and derive conditions the vacuum must satisfy. In Sec.~\ref{Main}, we discuss the relation of the gravitino masses to the gauge couplings and the prepotential, taking into account the vacuum conditions. Section~\ref{Summary} is devoted to the summary. In Appendix~\ref{notation}, we collect the spinor notations.

\section{Setup}\label{setup}
In this section, we specify the model. Here we follow the convention of Ref.~\cite{Trigiante:2016mnt}, and use the unit $M_{P}= 1$, where $M_{P}= 2.4\times 10^{18}$ GeV is the reduced Planck mass. We  introduce the only relevant parts of $\mathcal{N}=2$ supergravity for our purpose, and refer the literature~\cite{Andrianopoli:1996cm,DallAgata:2003sjo,DAuria:2004yjt,Andrianopoli:2011zj,Trigiante:2016mnt} for further details.

\subsection{Vector and hyper sectors}
The contents are given as follows:
\begin{align}
&{\rm{Vector\ multiplets:}}\ \ \{z^i, \lambda^{iA}, A^i_{\mu}\},\ \ (i=1,\cdots ,n_v)\\
&{\rm{Hypermultiplet:}}\ \ \{b^u,\zeta_{\alpha}\},\\
&{\rm{Gravitational\ multiplet:}}\ \ \{g_{\mu \nu}, \psi^A_{\mu}, A^0_{\mu}\}.
\end{align}
An Abelian vector multiplet contains a complex scalar $z^i$, two gauginos $\lambda^{iA}\ (A=1,2)$ and a vector $A^i_{\mu}$. Here the index $i$ labels the vector multiplets $(i=1,\cdots ,n_v)$. A hypermultiplet contains four real scalars $b^u\ (u=0,\cdots,3)$ and two hyperinos $\zeta_{\alpha}\ (\alpha=1,2)$.\footnote{Note that $\alpha$ is not a spinor index. The spinor indices are suppressed throughout this paper.} The gravitational multiplet contains the spacetime metric $g_{\mu \nu}\ (\mu,\nu=0,\cdots,3)$, two gravitinos $\psi^A_{\mu}\ (A=1,2)$ and the graviphoton $A_{\mu}^0$. 
Note that there are totally $n_v+1$ vector fields in the system and they are labeled by $A_{\mu}^{\Lambda}\ (\Lambda=0,1,\cdots, n_v)$.

\subsubsection*{Vector sector} 
The vector sector is governed by the prepotential $F(X^{\Lambda})$, which is a holomorphic and homogeneous function of degree two with $n_v+1$ complex variables $X^{\Lambda}\ (\Lambda=0,1,\cdots, n_v)$. In general, it can be parametrized as 
\begin{align}
F=-i(X^0)^2f(X^{i}/X^0),
\end{align}
where $f$ is an arbitrary holomorphic function. 
It is useful to define the following holomorphic section,
\begin{align}
\Omega ^M(z)=\left( \begin{array}{cc} X^{\Lambda }(z)\\ F_{\Sigma }(z)\\ \end{array} \right) , \ \ (\Lambda ,\Sigma =0,1,\cdots ,n_v)
\end{align}
where $F_{\Sigma}=\partial F/\partial X^{\Sigma}$, since the electric-magnetic duality that is a symmetry of $\mathcal{N}=2$ supergravity acts on the section. Note that $M$ labels $2n_v+2$ components.

Based on $\Omega$, the K\"ahler potential $\mathcal{K}$ is given by
\begin{align}
\mathcal{K}=-\log (i \bar{\Omega}^T\mathbb{C}\Omega)=-\log \left(i  \bar{X}^{\Lambda }F_{\Lambda }-i\bar{F}_{\Lambda }X^{\Lambda }\right),
\end{align}
where $\mathbb{C}$ is a symplectic invariant tensor, 
\begin{align}
\mathbb{C}=\begin{pmatrix} 
{\bm{0}}_{n_v+1} & {\bm{1}}_{n_v+1} \\ -{\bm{1}}_{n_v+1}& {\bm{0}}_{n_v+1}\\ 
\end{pmatrix}.
\end{align}

We take a special coordinate as
\begin{align}
X^0=1, \ \ X^i=z^i,
\end{align}
where $z^i $ are identified as physical scalars in the vector multiplets.
Then, the K\"ahler potential is written by
\begin{align}
\mathcal{K}=-\log \mathcal{K}_0,\ \ {\rm{where}} \ \ \mathcal{K}_0\equiv 2(f+\bar{f})-(z-\bar{z})^i(f_i-\bar{f}_i),
\end{align}
where the subscript $i$ on $f$ denotes the derivative with respect to $z^i$.

Finally, for later convenience, we list several quantities which appear in the Lagrangian and the supersymmetry transformations:
\begin{align}
&V^M\equiv e^{\mathcal{K}/2}\Omega^M = e^{\mathcal{K}/2}\left( \begin{array}{cc} X^{\Lambda }(z)\\ F_{\Sigma }(z)\\ \end{array} \right) ,\\
&U^M_i\equiv \nabla_i V^M=\left(  \partial_i +\frac{1}{2}\partial_i\mathcal{K}\right) V^M,\\
&\nabla_i U_j^M=\partial_iU_j^M +\frac{1}{2}\partial_i\mathcal{K} U_j^M- \Gamma ^k_{ij}U_k^M= e^{\mathcal{K}}f_{ijk}g^{k\bar{k}}\bar{U}_{\bar{k}}^M. \label{formula_1}
\end{align}

\subsubsection*{Hyper sector} 
As for the hyper sector, we consider the following metric~\cite{Ferrara:1995gu,Ferrara:1995xi},
\begin{align}
h_{uv}=\frac{1}{2(b^0)^2}\delta_{uv},\label{Q_metric}
\end{align}
which describes a nonlinear sigma model on SO$(4,1)/{\rm{SO}}(4)$. The vielbein $\mathcal{U}^{\alpha A}=\mathcal{U}^{\alpha A}_udb^u$ can be read off as
\begin{align}
\mathcal{U}^{\alpha A}=\frac{1}{2b^0}\epsilon^{\alpha \beta}\left(db^0-i\sum_{x=1}^3\tau ^x d b^x\right)_{\beta}^{\ \ A}, \label{vielbein}
\end{align}
where $A =1,2$ and $\alpha =1,2$ represent the SU(2) and Sp(2) indices respectively (their conventions are shown in Appendix~\ref{notation}). $\tau^x$ is the standard Pauli matrices.

Note that Eq.~$\eqref{Q_metric}$ depends only on $b^0$, but not $b^{1,2,3}$, which means there are three commuting isometries:
\begin{align}
b^m\rightarrow b^m+c^m,\ \ (m=1,2,3)\label{shift}
\end{align}
where $c^m$ are real constants. Then, the associated Killing vectors $k_m^u$ and the moment maps $\mathcal{P}_m^x$ are given by
\begin{align}
k_m^u=\delta_m^u,\ \ \mathcal{P}_m^x=\frac{1}{b^0}\delta_m^x. \label{killingmap}
\end{align}

\subsection{Gauging by embedding tensor} 
Now we consider to gauge the isometries~$\eqref{shift}$. For this purpose, we employ the embedding tensor formalism~\cite{deWit:2002vt, deWit:2005ub}, 
which is useful for discussing the general gauging of the extended supergravity. This formalism formally introduces a double copy of the gauge fields, i.e., the electric gauge fields~$A_\mu^\Lambda$ 
and the magnetic gauge fields~$A_{\mu\Sigma}$ $(\Lambda,\Sigma=0,1,\cdots,n_v)$, 
and gauges some of the global symmetries with the gauge couplings, 
\begin{align}
\Theta_{M}^{\ m}=\left( \begin{array}{cc} \Theta_{\Lambda }^{\ m}\\ \Theta^{\Sigma m}\\ \end{array} \right)=\left(\begin{array}{ccc} \Theta_{\Lambda}^{\ 1}& \Theta_{\Lambda}^{\ 2}&\Theta_{\Lambda}^{\ 3} \\ \Theta^{\Sigma 1}& \Theta^{\Sigma 2}& \Theta^{\Sigma 3}\\ \end{array} \right), \label{dET}
\end{align}
which are called the embedding tensor. In the following, we call $\Theta_{\Lambda }^{\ m}$ and $\Theta^{\Sigma m}$ as electric and magnetic couplings, respectively.

The tensor $\Theta_{M}^{\ m}$ must satisfy several conditions for the self-consistency of the theory~\cite{deWit:2002vt, deWit:2005ub}. In our case where no isometry on the vector sector is gauged, the only corresponding constraint is 
\begin{align}
\Theta_{M}^{\ m}\mathbb{C}^{MN}\Theta_{N}^{\ n}=0,\label{cond_theta}
\end{align}
or
\begin{align}
\Theta_{\Lambda}^{\ 1}\Theta^{\Lambda 2}-\Theta_{\Lambda}^{\ 2}\Theta^{\Lambda 1}=0,\label{ETconst12}\\
\Theta_{\Lambda}^{\ 2}\Theta^{\Lambda 3}-\Theta_{\Lambda}^{\ 3}\Theta^{\Lambda 2}=0,\label{ETconst23}\\
\Theta_{\Lambda}^{\ 3}\Theta^{\Lambda 1}-\Theta_{\Lambda}^{\ 1}\Theta^{\Lambda 3}=0.\label{ETconst31}
\end{align}
Then the covariant derivative is defined by
\begin{align}
D_{\mu}\equiv&\;
\partial_{\mu}-A_{\mu}^{\Lambda}\Theta_{\Lambda}^{\ m}T_m-A_{\mu\Sigma}\Theta^{\Sigma m}T_m, \label{def_covder}
\end{align}
where $T_m$ are generators of the isometries $\eqref{shift}$, thus $k_m^u=T_mb^u=\delta_m^u$. Note that the magnetic vectors $A_{\mu\Sigma}$ also participate in the gauging with the magnetic couplings~$\Theta^{\Sigma m}$. We also define 
\begin{align}
k_M^u=\Theta_{M}^{\ m}k_m^u,\ \ \mathcal{P}_M^x=\Theta_{M}^{\ m}\mathcal{P}_m^x. \label{KillingMap}
\end{align}
The introduction of the magnetic vector fields leads to the wrong counting of degree of freedom. In order to address the problem, we have to introduce two-form auxiliary fields, which enlarge gauge symmetries, and then, modify the kinetic terms for vector fields and add topological couplings accordingly. As these couplings do not affect to the following discussion, we do not write their explicit forms (see\cite{Andrianopoli:1996cm,DallAgata:2003sjo,DAuria:2004yjt,Andrianopoli:2011zj,Trigiante:2016mnt} for the whole expressions).

\subsection{Supersymmetry transformation}
Here we show the supersymmetry transformations of the fermions, which are necessary to discuss the supersymmetry breaking conditions (and patterns) in the next section.  
The relevant parts of supersymmetry transformations are given by~\cite{Trigiante:2016mnt},
\begin{align}
&\delta \psi_{A\mu}=i\mathbb{S}_{AB}\gamma_{\mu}\epsilon^B+\cdots,\label{delpsi}\\
&\delta \lambda^{\bar{i}}_{A}=\bar{W}^{\bar{i}}_{AB}\epsilon^B+\cdots,\\
&\delta \zeta^{\alpha}=\bar{N}^{\alpha}_{\ A}\epsilon^A+\cdots,\label{delzeta}
\end{align}
where 
\begin{align}
&\mathbb{S}_{AB}\equiv \frac{i}{2}(\tau ^x)_{AB}\mathcal{P}_M^xV^M,\label{psipsi}\\
&W^{iAB}\equiv i(\tau ^x)^{AB}\mathcal{P}_M^xg^{i\bar{j}}\bar{U}_{\bar{j}}^M,\\
&N_{\alpha }^A\equiv -2\mathcal{U}^A_{u \alpha }k_M^u\bar{V}^M,    
\end{align}
and $\bar{W}^{\bar{i}}_{AB}\equiv (W^{iAB})^*, \bar{N}^{\alpha}_{\ A}\equiv (N_{\alpha}^{\ A})^*$. The ellipses in Eqs.~$\eqref{delpsi}$-$\eqref{delzeta}$ represent terms which vanish in the Minkowski background. The matrices are given explicitly by
\begin{align}
&\mathbb{S}_{AB} =  \frac{-ie^{\mathcal{K}/2}}{2b^0} \left(\begin{array}{cc}  i\beta-\alpha&\gamma \\
\gamma& i\beta+\alpha\\ \end{array} \right)  ,\label{S}\\    
&\bar{W}^{\bar{i}}_{AB} = -\frac{ie^{\mathcal{K}/2}}{b^0}g^{i\bar{i}} \left(\begin{array}{cc}   \nabla_i(i\beta-\alpha)&\nabla_i\gamma \\
\nabla_i\gamma& \nabla_i(i\beta+\alpha)\\\end{array} \right)  ,\label{W}\\
&\bar{N}^{\alpha}_{\ A} =  \frac{ie^{\mathcal{K}/2}}{b^0} \left(\begin{array}{cc}   \gamma&i\beta+\alpha \\
-i\beta+\alpha& -\gamma\\\end{array} \right) .\label{N}
\end{align}
Here $\alpha, \beta$ and $\gamma$ are defined by 
\begin{align}
&\alpha \equiv \Theta_{M}^{\ 1}\Omega^M =(\Theta_{\Lambda}^{\ 1}X^{\Lambda}+\Theta^{\Lambda 1}F_{\Lambda}),\label{defx}\\
&\beta \equiv \Theta_{M}^{\ 2}\Omega^M =(\Theta_{\Lambda}^{\ 2}X^{\Lambda}+\Theta^{\Lambda 2}F_{\Lambda}),\label{defy}\\
&\gamma \equiv \Theta_{M}^{\ 3}\Omega^M =(\Theta_{\Lambda}^{\ 3}X^{\Lambda}+\Theta^{\Lambda 3}F_{\Lambda}),\label{defz}
\end{align}
and we introduced their covariant derivatives as 
\begin{align}
\nabla_i \alpha =\partial_i\alpha+\partial_i\mathcal{K}\alpha=e^{-\mathcal{K}/2}\Theta_{M}^{\ 1}U_i^M,
\end{align}
and so on. 


\section{Gravitino masses}\label{GM}
To discuss the supersymmetry breaking patterns, we need to identify the order parameters of the supersymmetry breaking. In the global supersymmetric theory, the goldstino(s) appears if the supersymmetry is spontaneously broken, and the order parameter (or breaking scale) can be read off from the goldstino transformations. In supergravity, the goldstino(s) is absorbed by the gravitino(s) through the super-higgs mechanism, and the gravitino(s) acquires a mass, which is related to the supersymmetry breaking scale at the vacuum. Therefore, in this section, we derive the expressions of the gravitino masses under the general gauging. 

\subsection{Gravitino masses and goldstino transformations}     
The corresponding parts in $\mathcal{N}=2$ supergravity Lagrangian are given by~\cite{Trigiante:2016mnt},
\begin{align}
\nonumber \mathcal{L}=&2\mathbb{S}_{AB}\bar{\psi}_{\mu}^A\gamma^{\mu\nu}\psi_{\nu}^B+ig_{i\bar{j}}\bar{W}^{\bar{j}}_{AB}\bar{\lambda}^{iA}\gamma_{\mu}\psi^{\mu B}+2i\bar{N}^{\alpha}_{\ A}\bar{\zeta}_{\alpha}\gamma_{\mu}\psi^{\mu A}+{\rm{h.c.}},\\
=&-\frac{ie^{\mathcal{K}/2}}{b^0}(\bar{\psi}_{\mu}^1,\bar{\psi}_{\mu}^2)M_{\psi}\gamma^{\mu\nu}\left( \begin{array}{cc} \psi_{\nu}^1\\ \psi_{\nu}^2\\ \end{array} \right)+\frac{e^{\mathcal{K}/2}}{b^0}(\bar{\chi}_1\gamma_{\mu}\psi^{\mu 1}+\bar{\chi}_2\gamma_{\mu}\psi^{\mu 2})+{\rm{h.c.}}, \label{GG}
\end{align}
where 
\begin{align}
M_{\psi}=\left(\begin{array}{cc}  i\beta-\alpha&\gamma \\
\gamma& i\beta+\alpha\\ \end{array} \right),
\end{align}
and we defined the goldstinos as
\begin{align}
&\chi_1=2(i\beta-\alpha)\zeta_2-2\gamma\zeta_1+\nabla_i(i\beta-\alpha) \lambda^{i1}+\nabla_i\gamma \lambda^{i2},\label{chi1}\\
&\chi_2=2\gamma\zeta_2-2(i\beta+\alpha)\zeta_1+\nabla_i\gamma \lambda^{i1}+\nabla_i(i\beta+\alpha) \lambda^{i2}.\label{chi2}
\end{align}

Then, we need to diagonalize their mass matrix $M_{\psi}$. This can be achieved by a unitary matrix $U$,
\begin{align}
U^{T}M_{\psi}U=\left(\begin{array}{cc}  \sigma_1&0 \\
0& \sigma_2\\ \end{array} \right), \ \ \sigma_2\geq\sigma_1\geq 0
\end{align}
where $\sigma_A (A=1,2)$ are the singular values of $M_{\psi}$, and given explicitly by 
\begin{align}
&\sigma _{1}=X_+-X_-,\ \ \sigma _{2}=X_++X_-,\label{sigma12}\\
&X_{\pm}\equiv \frac{1}{\sqrt{2}}\sqrt{|\alpha|^2+|\beta|^2+|\gamma|^2\pm |\alpha^2+\beta^2+\gamma^2|}.\label{Xpm}
\end{align}
Defining new gravitinos and goldstinos by  
\begin{align}
\left( \begin{array}{cc} \tilde{\psi}_1\\ \tilde{\psi}_2\\ \end{array} \right)=U^T\left( \begin{array}{cc} \psi_1\\ \psi_2\\ \end{array} \right),\ \ \left( \begin{array}{cc} \tilde{\chi}_1\\ \tilde{\chi}_2\\ \end{array} \right)=\left(\begin{array}{cc}  1/\sigma_1&0 \\
0& 1/\sigma_2\\ \end{array} \right)U^T\left( \begin{array}{cc} \chi_1\\ \chi_2\\ \end{array} \right),\label{new_psichi}
\end{align}
we can rewrite the Lagrangian~$\eqref{GG}$ as
\begin{align}
\mathcal{L}=-\frac{ie^{\mathcal{K}/2}}{b^0}\sum_{A=1,2}\sigma_A\left(\bar{\tilde{\psi}}^A_{\mu}\gamma^{\mu\nu}\tilde{\psi}^A_{\nu}+i\bar{\tilde{\chi}}_A\gamma_{\mu}\tilde{\psi}^{A\mu}\right)+{\rm{h.c.}}.\label{tilde_GG}
\end{align}

Let us show that the transformations of the goldstinos defined in Eq.~$\eqref{new_psichi}$ are also characterized by $\sigma_A$. First, using Eqs.~$\eqref{delpsi}$-$\eqref{delzeta}$, the supersymmetry transformations of $\chi_1$ and $\chi_2$ are evaluated as
\begin{align}
\left( \begin{array}{cc} \delta\chi_1\\ \delta\chi_2\\ \end{array} \right) = \frac{ie^{\mathcal{K}/2}}{b^0} M_{\chi}   \left( \begin{array}{cc} \epsilon _1\\ \epsilon _2\\ \end{array} \right),\label{Tchi}
\end{align}
where $M_{\chi}$ is a $2\times 2$ hermitian matrix whose components are 
\begin{align}
\nonumber M_{\chi11}=& 2(|\alpha|^2+|\beta|^2+|\gamma|^2)-4{\rm{Im}}(\alpha\bar{\beta})\\
&+|\nabla \alpha|^2+|\nabla \beta|^2+|\nabla \gamma|^2-2{\rm{Im}}(\nabla \alpha\cdot \overline{\nabla \beta}),\\
M_{\chi12}=&M_{\chi21}^*=-4{\rm{Im}}(\beta\bar{\gamma})-4i{\rm{Im}}(\alpha\bar{\gamma})-2{\rm{Im}}(\nabla \beta\cdot \overline{\nabla \gamma})-2i{\rm{Im}}(\nabla \alpha\cdot \overline{\nabla \gamma}),\\
\nonumber M_{\chi22}=&2(|\alpha|^2+|\beta|^2+|\gamma|^2)+4{\rm{Im}}(\alpha\bar{\beta})\\
&+|\nabla \alpha|^2+|\nabla \beta|^2+|\nabla \gamma|^2+2{\rm{Im}}(\nabla \alpha\cdot \overline{\nabla \beta}).
\end{align}
Here we have introduced the notation, $\nabla \alpha \cdot \overline{\nabla \beta} =g^{i\bar{j}}\nabla_i\alpha\bar{\nabla}_{\bar{j}}\bar{\beta}$ and $|\nabla \alpha|^2  =\nabla \alpha \cdot \overline{\nabla \alpha}$. 
This inner product is positive definite, that is, $|\nabla \alpha|^2\geq 0$ and $|\nabla \alpha|^2= 0$ if and only if $\nabla_i \alpha= 0$.
Next, from the supergravity Ward identity,
\begin{align}
\delta_B^AV=-12\bar{\mathbb{S}}^{AC}\mathbb{S}_{BC}+g_{i\bar{j}}W^{iAC}\bar{W}^{\bar{j}}_{BC}+2N_{\alpha}^{\ A}\bar{N}^{\alpha}_{\ B},
\end{align}
where $\bar{\mathbb{S}}^{AB}=(\mathbb{S}_{AB})^*$, we obtain the expression of the scalar potential and three equations:
\begin{align}
&V=\frac{e^{\mathcal{K}}}{(b^0)^2}(-|\alpha|^2-|\beta|^2-|\gamma|^2+|\nabla \alpha|^2+|\nabla \beta|^2+|\nabla \gamma|^2),\label{V}\\
&0={\rm{Im}}( \alpha \bar{ \beta})-{\rm{Im}}(\nabla \alpha\cdot \overline{\nabla \beta}),\label{W1}\\
&0={\rm{Im}}( \beta \bar{\gamma})-{\rm{Im}}(\nabla \beta\cdot \overline{\nabla \gamma}),\label{W2}\\
&0={\rm{Im}}( \gamma \bar{\alpha})-{\rm{Im}}(\nabla \gamma\cdot \overline{\nabla \alpha}).\label{W3}
\end{align}
In this paper, we focus on the Minkowski vacuum, and therefore, the following equation   
\begin{align}
0=-|\alpha|^2-|\beta|^2-|\gamma|^2+|\nabla \alpha|^2+|\nabla \beta|^2+|\nabla \gamma|^2, \label{Mcond}
\end{align}
is satisfied at the vacuum. From Eqs.~$\eqref{W1}$-$\eqref{W3}$, and~$\eqref{Mcond}$, we can rewrite $M_{\chi}$ as   
\begin{align}
&M_{\chi11}= 3(|\alpha|^2+|\beta|^2+|\gamma|^2)-6{\rm{Im}}(\alpha\bar{\beta}),\\
&M_{\chi12}=M_{21}^*=-6{\rm{Im}}(\beta\bar{\gamma})-6i{\rm{Im}}(\alpha\bar{\gamma}),\\
&M_{\chi22}=3(|\alpha|^2+|\beta|^2+|\gamma|^2)+6{\rm{Im}}(\alpha\bar{\beta}).
\end{align}
Thus, we obtain a relation
\begin{align}
M_{\chi}=3M_{\psi}M_{\psi}^{\dagger}.
\end{align}
By definition, the unitary matrix $U$ satisfies
\begin{align}
U^TM_{\psi}M_{\psi}^{\dagger}U^*=\left(\begin{array}{cc}  \sigma_1^2&0 \\
0& \sigma_2^2\\ \end{array} \right),
\end{align}
which implies that we can take $U$ as a matrix diagonalizing the supersymmetry transformation of the goldstinos~$\eqref{Tchi}$, and obtain
\begin{align}
\left( \begin{array}{cc} \delta\tilde{\chi}_1\\ \delta\tilde{\chi}_2\\ \end{array} \right) = \frac{3ie^{\mathcal{K}/2}}{b^0}\left(\begin{array}{cc}  \sigma_1&0 \\
0& \sigma_2\\ \end{array} \right)   \left( \begin{array}{cc} \tilde{\epsilon} _1\\ \tilde{\epsilon} _2\\ \end{array} \right),\ \ \left( \begin{array}{cc} \tilde{\epsilon}_1\\ \tilde{\epsilon}_2\\ \end{array} \right)\equiv U^T\left( \begin{array}{cc} \epsilon_1\\ \epsilon_2\\ \end{array} \right). \label{T_tilde_chi}
\end{align}
As also understood from this expressions, $\sigma_1$ and $\sigma_2$ characterize the supersymmetry breaking as they should.

Let us go back to the Lagrangian~$\eqref{tilde_GG}$. As is obvious from Eq.~$\eqref{T_tilde_chi}$, the goldstinos are eliminated by taking a unitary gauge,
\begin{align}
\tilde{\chi}_{A}=0,    
\end{align}
and we obtain the canonical gravitino masses
\begin{align}
m_{A}=\frac{e^{\mathcal{K}/2}}{b^0}\sigma_A,\ \ A=1,2.    
\end{align}
In the following, we focus on the behaviours of $\sigma_A$ by neglecting a common factor $\frac{e^{\mathcal{K}/2}}{b^0}$.\footnote{We regard $\sigma_A$ as dimensionless quantities. Therefore, the gravitino masses are given by $m_{A}=\frac{e^{\mathcal{K}/2M_{P}^2}}{b^0}M_{P}^2\sigma_A$ when the Planck scale is recovered.}


\subsection{Behaviours of gravitino masses at first sight} 

\begin{figure}[t]
\hspace*{4cm}
\includegraphics[width=8.0cm]{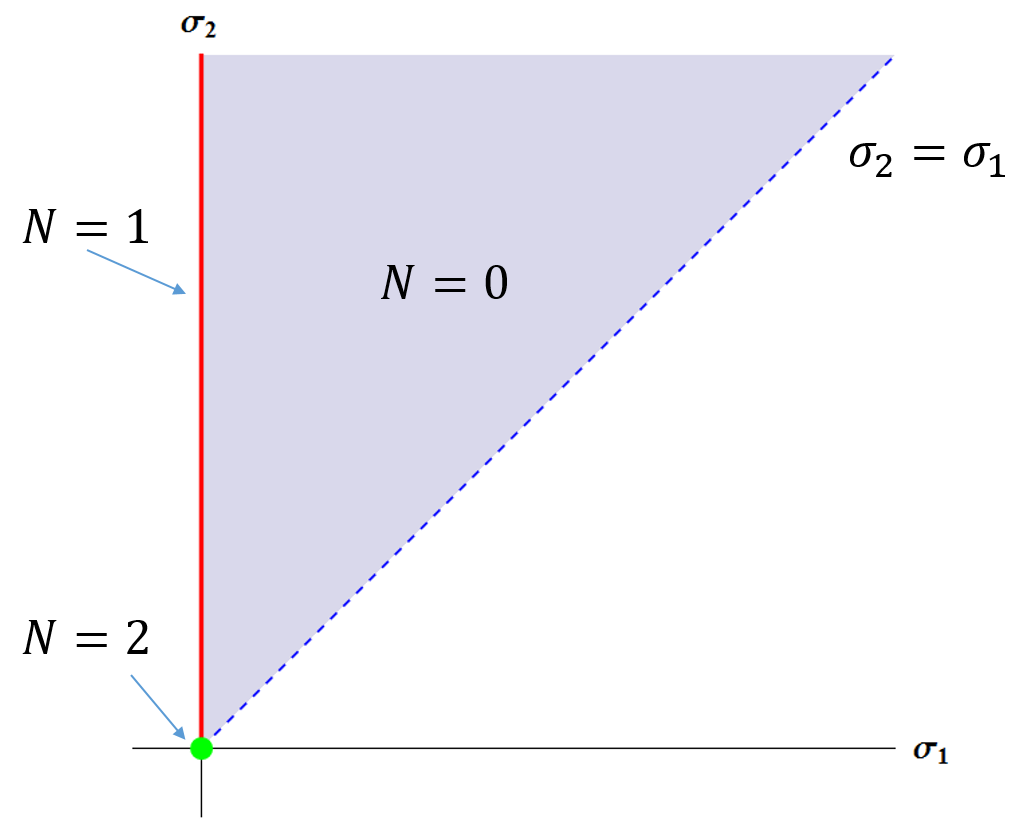}
\caption{$(\sigma_1$-$ \sigma_2)$ plane. The $\mathcal{N}=2$ preserving vacuum is located at the origin, and $\mathcal{N}=1$ preserving (or partially broken) vacuum is on the line $\sigma_1=0$ with $\sigma_2\neq 0$. The other region corresponds to the $\mathcal{N}=0$ (fully broken) vacua.}
\label{region}
\end{figure}

By definition, $(\sigma_1$-$ \sigma_2)$ plane has a domain which is restricted by $\sigma_2\geq \sigma_1$ (Fig.~\ref{region}). 
Obviously, an $\mathcal{N}=2$ preserving vacuum is located at its origin, and $\mathcal{N}=1$ preserving (or partially broken) vacuum corresponds to the vertical axis of $\sigma_1=0$ with $\sigma_2\neq 0$. The other region corresponds to the $\mathcal{N}=0$ (fully broken) vacua.

\begin{figure}[t]
\hspace*{4cm}
\includegraphics[width=8.0cm]{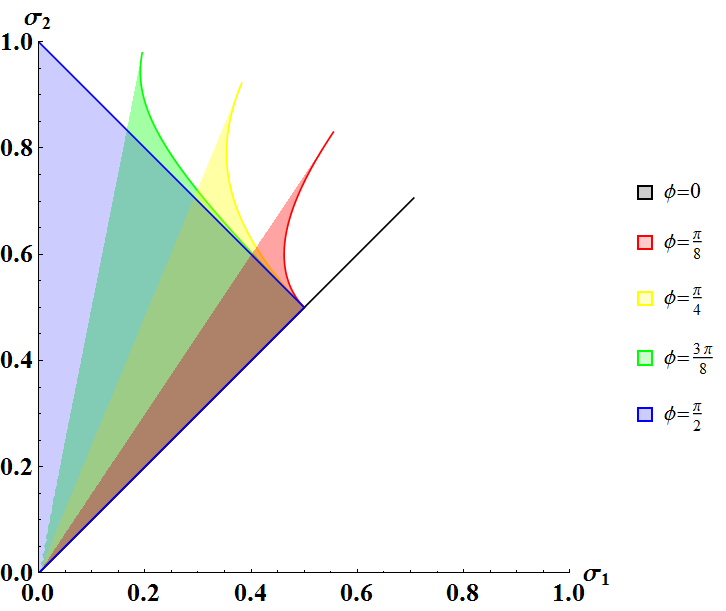}
\caption{$\phi\equiv {\rm{arg}}\alpha-{\rm{arg}}\beta$ dependence of $\sigma_1$ and $\sigma_2$ in the case with two isometries gauged. As the parameters~$|\alpha|$ and $|\beta|$, we change them under $0\leq|\alpha|,|\beta|\leq 0.5$.}
\label{phi-dependence2}
\end{figure}

Let us comment on the relation between the number of gaugings and the breaking patterns. As can be seen from the expressions~$\eqref{sigma12}$ and~$\eqref{Xpm}$, when only one isometry is gauged, e.g., $\alpha \neq 0$ and $\beta =\gamma = 0$, we always have the degenerate breaking scales
\begin{align}
\sigma_1=\sigma_2=|\alpha|.
\end{align}
On the other hand, for the case with the two directions gauged, e.g., $\alpha,\beta \neq 0$ and $\gamma = 0$, we have rich breaking patterns. In this case, we can parametrize $\sigma_A (A=1,2)$ as
\begin{align}
&\sigma_1=X_+-X_-, \ \ \sigma_2=X_++X_-, \\
&X_{\pm}=\frac{1}{\sqrt{2}}\sqrt{|\alpha|^2+|\beta|^2\pm \sqrt{|\alpha|^4+|\beta|^4+2|\alpha|^2|\beta|^2{\rm{cos}}2\phi}},
\end{align}
where $\phi\equiv {\rm{arg}}\alpha-{\rm{arg}}\beta$. The figure~\ref{phi-dependence2} shows the parametric plot in $(\sigma_1$-$ \sigma_2)$ plane under $0\leq|\alpha|,|\beta|\leq 0.5$ with fixed $\phi=\{ 0,\frac{\pi}{8},\frac{\pi}{4},\frac{3\pi}{8},\frac{\pi}{2}\}$.
When $\phi=0$, we have degenerate breaking scales. As $\phi$ approaches to $\pi /2$, the breaking scales can be hierarchical and the $\mathcal{N}=1$ (partial breaking) can be covered when $\phi=\pi /2$. 

To conclude, we have obtained general expressions of gravitino masses (or supersymmetry breaking scales) in the model containing a single hypermultiplet gauged by $n_v$-Abelian vector multiplets and graviphoton. These observations are based on the assumption that the parameters $\Theta_M^{\ m}$ and the prepotential $f$ (or their specific combinations $\alpha, \beta,$ and $\gamma$) can be changed independently. We need to check that these breaking patterns are really realized at the minimum since the minimization conditions of the scalar potential should impose some constraints between the parameters and vacuum expectation values of $z^i$, which is going to be a topic in the next section.

\subsection{Condition for special cases} 
Before going to the detailed analysis of the scalar potential, let us discuss special cases, where $\mathcal{N}=2$ and $1$ supersymmetries are preserved. At these vacua, the parameters $\Theta_M^{\ m}$ or $\alpha,\beta,\gamma$ are further restricted by several conditions. Here we summarize them for later convenience.
The following discussion is based on the approach of Ref.~\cite{Louis:2009xd}.

\subsubsection{$\mathcal{N}=2$ (no breaking)}\label{N=2_cond}
From Eq.~$\eqref{sigma12}$, the condition for $\mathcal{N}=2$ preserving vacuum is $X_+=X_-=0$, which leads to 
\begin{align}
\alpha =\beta =\gamma =0.\label{n=2cond}
\end{align}
In terms of the embedding tensor, these equations can be written as  
\begin{align}
\Theta_{\Lambda}^{\ m}X^{\Lambda}+\Theta^{\Lambda m}F_{\Lambda}=0, \ \ (m=1,2,3) \label{n=2condtheta}
\end{align}
Also, from Eq.~$\eqref{Mcond}$, we have  
\begin{align}
\nabla_i\alpha=\nabla_i\beta =\nabla_i\gamma=0.\label{n=2cond2}
\end{align}
Under the conditions~$\eqref{n=2cond}$, the equations~$\eqref{n=2cond2}$ give 
\begin{align}
\Theta_i^{\ m}+F_{i\Lambda}\Theta^{\Lambda m}=0,\ \ (m=1,2,3) \label{n=2cond2theta}
\end{align}
since $\partial_iX^{\Lambda}=\delta^{\Lambda}_i$ and $\partial_iF_{\Lambda}=F_{\Lambda i}$ in the special coordinate. Multiplying $z^i$ to Eq.~$\eqref{n=2cond2theta}$ and subtracting Eq.~$\eqref{n=2condtheta}$, we obtain
\begin{align}
\Theta_{0}^{\ m}+F_{0 \Lambda}\Theta^{\Lambda m}=0, \label{0cond}
\end{align}
where we have used the property $F_{\Lambda}=F_{\Lambda \Sigma}X^{\Sigma}$. As a result, Eqs.~$\eqref{n=2cond2theta}$ and~$\eqref{0cond}$ are summarized as  
\begin{align}
\Theta_{\Lambda}^{\ m}+F_{\Lambda\Sigma}\Theta^{\Sigma m}=0. \label{theta_vanish}
\end{align}
Since the matrix ${\rm{Im}}F_{\Lambda\Sigma}$ has to be invertible for special geometry, the equation~$\eqref{theta_vanish}$ leads to 
\begin{align}
\Theta_{M}^{\ m}=0, 
\end{align}
which means that no gauging is a solution in our setup. 


\subsubsection{$\mathcal{N}=1$ (partial breaking)} \label{N=1}
Next, we derive the conditions for the partial breaking. Obviously, it occurs when $X_+=X_-$, that is,
\begin{align}
\alpha^2+\beta^2+\gamma^2=0. \label{PB_cond}
\end{align}
In the following, we see the consequence of this equation, dividing the cases by the number of gaugings. 

\subsubsection*{$(i)$ Gauging one direction}
Let us consider a case, $\Theta_M^{\ 1}\neq 0$ and $\Theta_M^{\ 2,3}=0$. Then, the $\mathcal{N}=1$ preserving condition~$\eqref{PB_cond}$ and Eq.~$\eqref{Mcond}$ imply
\begin{align}
\alpha=\nabla_i \alpha=0. 
\end{align}
In the same way with the subsection~\ref{N=2_cond}, we obtain
\begin{align}
\Theta_{M}^{\ 1}=0, 
\end{align}
which contradicts with $\Theta_M^{\ 1}\neq 0$. Therefore, the one isometry gauging cannot realize the partial breaking.

\subsubsection*{$(ii)$ Gauging two directions}
Next, we assume that $\Theta_M^{\ 1,2}\neq 0$ and $\Theta_M^{\ 3}= 0$. The condition~$\eqref{PB_cond}$ requires either of $\alpha \pm i\beta =0$. Then, the equations.~$\eqref{W1}$ and $\eqref{Mcond}$ read
\begin{align}
&2|\alpha|^2=|\nabla\alpha|^2+|\nabla\beta|^2,\\
&\mp |\alpha|^2={\rm{Im}}(\nabla \alpha \cdot \overline{\nabla \beta }). 
\end{align}
By summing these two equations, we obtain $|\nabla\alpha \pm i\nabla\beta|^2=0$. Therefore, we also have $\nabla_i\alpha \pm i\nabla_i\beta=0$. By repeating the same process, we obtain  
\begin{align}
\Theta_{\Lambda}^{\ 1}+F_{\Lambda\Sigma}\Theta^{\Sigma 1}\pm i\left(\Theta_{\Lambda}^{\ 2}+F_{\Lambda\Sigma}\Theta^{\Sigma 2}\right) =0.\label{sol_PB2}
\end{align}
Note that if there is no magnetic couplings, i.e., $\Theta^{\Sigma 1,2}=0$, Eq.~$\eqref{sol_PB2}$ leads to $\Theta_{\Lambda}^{\ 1,2}=0$, which contradicts to the assumption~$\Theta_M^{\ 1,2}\neq 0$. Therefore, the introduction of the magnetic coupling is necessary for Eq.~$\eqref{sol_PB2}$ to have solutions.     
 

\subsubsection*{$(iii)$ Gauging three directions}
Finally, we consider the case with $\Theta_M^{\ 1,2,3}\neq 0$. The solution of Eq.~$\eqref{PB_cond}$ can be parametrized by
\begin{align}
\alpha+i\beta=w\gamma, \ \ \alpha-i\beta=-\frac{1}{w}\gamma,
\end{align}
with a non-vanishing complex number $w$. Then, from Eq.~$\eqref{Mcond}$ and Eqs.~$\eqref{W1}$-$\eqref{W3}$, we obtain
\begin{align}
&\left(|w|+\frac{1}{|w|}\right)^2|\gamma|^2=2(|\nabla \alpha|^2+|\nabla \beta|^2+|\nabla \gamma|^2),\\
&\left(|w|^2-\frac{1}{|w|^2}\right)|\gamma|^2=4{\rm{Im}}\nabla \alpha \cdot \overline{\nabla \beta },\\
&\left(w+\frac{1}{w}+{\rm{c.c.}}\right)|\gamma|^2=-4{\rm{Im}}\nabla \beta \cdot \overline{\nabla \gamma },\\
&\left(w-\frac{1}{w}-{\rm{c.c.}}\right)|\gamma|^2=-4i{\rm{Im}}\nabla \gamma \cdot \overline{\nabla \alpha }.
\end{align}
Based on these equations, it is straightforward to show that the following equation 
\begin{align}
|\nabla \alpha+i \nabla \beta-w\nabla \gamma|^2+|w|^2|\nabla \alpha-i\nabla \beta+\frac{1}{w}\nabla \gamma|^2=0,
\end{align}
holds, which implies
\begin{align}
\nabla_i\alpha+i\nabla_i\beta=w\nabla_i\gamma, \ \ \nabla_i\alpha-i\nabla_i\beta=-\frac{1}{w}\nabla_i \gamma. \label{PBcond3}
\end{align}
Note that $w=\pm 1$ and $w=\pm i$ imply $\Theta_M^{\ 1}= 0$ and $\Theta_M^{\ 2}= 0$ respectively, and we exclude these cases.
Then, in terms of the embedding tensor, we have 
\begin{align}
&\Theta_{\Lambda}^{\ 1}+F_{\Lambda\Sigma}\Theta^{\Sigma 1}+ i\left(\Theta_{\Lambda}^{\ 2}+F_{\Lambda\Sigma}\Theta^{\Sigma 2}\right) =w\left(\Theta_{\Lambda}^{\ 3}+F_{\Lambda\Sigma}\Theta^{\Sigma 3}\right),\label{1sol_PB3}\\
&\Theta_{\Lambda}^{\ 1}+F_{\Lambda\Sigma}\Theta^{\Sigma 1}- i\left(\Theta_{\Lambda}^{\ 2}+F_{\Lambda\Sigma}\Theta^{\Sigma 2}\right) =-\frac{1}{w}\left(\Theta_{\Lambda}^{\ 3}+F_{\Lambda\Sigma}\Theta^{\Sigma 3}\right),\label{2sol_PB3}
\end{align}
which give $4(n_v+1)$ equations. Since Eqs.~$\eqref{1sol_PB3}$ and~$\eqref{2sol_PB3}$ imply $w=0$ when $\Theta^{\Sigma 1,2,3}=0$, we can conclude that there is no solution in the absence of the magnetic couplings, also in this case.


\section{Scalar potential analysis}\label{SCA}
In this section, we discuss the scalar potential~$\eqref{V}$ and its minimum. Here we show it again:
\begin{align}
V=\frac{e^{\mathcal{K}}}{(b^0)^2}(-|\alpha|^2-|\beta|^2-|\gamma|^2+|\nabla \alpha|^2+|\nabla \beta|^2+|\nabla \gamma|^2). \label{V1} 
\end{align}
Note that it can be rewritten as
\begin{align}
V=r_{MN}(U^{MN}-V^M\bar{V}^N),
\end{align}
where
\begin{align}
&U^{MN}\equiv g^{i\bar{j}}U_i^M\bar{U}_{\bar{j}}^N,\ \  \bar{U}^{MN}=U^{NM},\\
&r_{MN}\equiv \frac{1}{(b^0)^2}\left(\sum_{m=1}^3 \Theta_M^{\ m}\Theta_N^{\ m}\right).
\end{align}

The stationary point of the scalar potential is given by solving  
\begin{align}
&\frac{\partial V}{\partial b^{0}}=0,\label{Vb_0}\\
&\frac{\partial V}{\partial z^i}=0.\label{Vz_i}
\end{align}
The former~$\eqref{Vb_0}$ is equivalent to impose 
\begin{align}
 -|\alpha|^2-|\beta|^2-|\gamma|^2+|\nabla \alpha|^2+|\nabla \beta|^2+|\nabla \gamma|^2=0,
\end{align}
which is already ensured by the Minkowski vacuum condition~$\eqref{Mcond}$. As for the latter~$\eqref{Vz_i}$, we can compute it as  
\begin{align}
\nonumber \frac{\partial V}{\partial z^i}=&r_{MN}(\partial_iU^{MN}-U_i^M\bar{V}^N)\\
\nonumber =&r_{MN}\left(g^{j\bar{k}}\bar{U}^N_{\bar{k}}(\partial_iU_j^{M}-\Gamma_{ij}^kU_k^M)+g^{j\bar{k}}U_j^M\partial_i\bar{U}_{\bar{k}}^N-U_i^M\bar{V}^N\right)\\
\nonumber =&r_{MN}g^{j\bar{k}}\bar{U}^N_{\bar{k}}\nabla_iU_j^M\\
=&r_{MN}e^{\mathcal{K}}\bar{U}^N_{\bar{j}}\bar{U}^M_{\bar{k}}g^{j\bar{j}}g^{k\bar{k}}f_{ijk}.
\end{align}
In the derivation, we have used $\partial_ig^{j\bar{k}}=-\Gamma_{ik}^jg^{k\bar{k}}$, $\partial_i\bar{U}_{\bar{k}}^N=\frac{1}{2}\partial_i\mathcal{K}\bar{U}_{\bar{k}}^N+g_{i\bar{k}}\bar{V}^N$, and Eq.~$\eqref{formula_1}$. Then, in terms of $\alpha,\beta,$ and $\gamma$, the equation~$\eqref{Vz_i}$ is summarized as 
\begin{align}
e^{2\mathcal{K}}(\bar{\nabla}_{\bar{j}}\bar{\alpha}\bar{\nabla}_{\bar{k}}\bar{\alpha}+\bar{\nabla}_{\bar{j}}\bar{\beta}\bar{\nabla}_{\bar{k}}\bar{\beta}+\bar{\nabla}_{\bar{j}}\bar{\gamma}\bar{\nabla}_{\bar{k}}\bar{\gamma})g^{j\bar{j}}g^{k\bar{k}}f_{ijk}=0.\label{Vi=0}
\end{align}

As summary, we derived the conditions the vacuum must satisfy: Eqs.~$\eqref{Mcond}$ and $\eqref{Vi=0}$. Also, the embedding tensor must satisfy the constraints~$\eqref{ETconst12}$-$\eqref{ETconst31}$. We need to investigate the behaviors of $\sigma_A$, under these conditions. In general, they depend on the values of gauge coupling constants $\Theta_M^{\ m}$ and the form of the prepotential or $f(z^i)$. Furthermore, $\Theta_M^{\ m}$ depends on the number of vector multiplets $(n_v)$ and gauging $(m=1,2,3)$. In the next section, therefore, we consider several concrete examples.

Before closing this section, let us comment on the three equations of the supergravity identity~$\eqref{W1}$-$\eqref{W3}$. Indeed, they are equivalent to the constraints on the embedding tensor~$\eqref{ETconst12}$-$\eqref{ETconst31}$. This can be seen by noting that the right-hand-side in Eqs.~$\eqref{W1}$-$\eqref{W3}$ can be expressed as   
\begin{align}
e^{-\mathcal{K}}\Theta_M^{\ m}\Theta_N^{\ n}{\rm{Im}}(V^M\bar{V}^N-U^{MN}), \ \ m\neq n, \ \  m,n=1,2,3.  \label{re_Ward}
\end{align}
Then, note the following relation~\cite{Andrianopoli:2015rpa},
\begin{align}
U^{MN}=-\frac{1}{2}\mathcal{M}^{MN}-\frac{i}{2}\mathbb{C}^{MN}-\bar{V}^MV^N,\label{Umn}
\end{align}
where $\mathcal{M}^{MN}$ is a symmetric matrix (see~\cite{Andrianopoli:2015rpa} for the explicit expression). By substituting Eq.~$\eqref{Umn}$ into Eq.~$\eqref{re_Ward}$, the remaining parts are 
\begin{align}
\frac{1}{2}e^{-\mathcal{K}}\Theta_M^{\ m}\Theta_N^{\ n}\mathbb{C}^{MN},
\end{align}
which vanish under the constraints~$\eqref{W1}$-$\eqref{W3}$.


\section{Behaviors of two supersymmetry breaking scales in explicit models}\label{Main}
In this section, we consider some examples which satisfy the different vacuum conditions, and investigate how the supersymmetry breaking scales change.    

\subsection{Single vector multiplet}
First, let us focus on the case of a single vector multiplet $(n_v=1)$.
In this case, the conditions~$\eqref{Mcond}$ and~$\eqref{Vi=0}$ become
\begin{align}
&-|\alpha|^2-|\beta|^2-|\gamma|^2+g^{z\bar{z}}|\nabla_z \alpha|^2+g^{z\bar{z}}|\nabla _z\beta|^2+g^{z\bar{z}}|\nabla_z \gamma|^2=0,\label{V=0_single}\\
&((\bar{\nabla}_{\bar{z}}\bar{\alpha})^2+(\bar{\nabla}_{\bar{z}}\bar{\beta})^2+(\bar{\nabla}_{\bar{z}}\bar{\gamma})^2)(g^{z\bar{z}})^2f_{zzz}=0.\label{Vi=0_single}
\end{align}
As for Eq.~$\eqref{Vi=0_single}$, we have two choices:
\begin{align}
{\rm{Case\ A:}}\ \ f_{zzz}=0.\label{V_i=01}
\end{align}
or
\begin{align}
{\rm{Case\ B:}}\ \ f_{zzz}\neq 0, \ \ {\rm{or}}\ \  (\nabla_z\alpha)^2+(\nabla_z\beta)^2+(\nabla_z\gamma)^2=0,\label{V_i=02}
\end{align}
since $g^{z\bar{z}}\neq 0$. Let us consider the two cases separately below.

\subsubsection{Case A}
Let us start from the case, $\eqref{V_i=01}$. Here, we assume that the prepotential takes the form
\begin{align}
f=z,    
\end{align}
which obviously satisfies $f_{zzz}=0$. Then, the vacuum condition~$\eqref{V=0_single}$ is reduced to 
\begin{align}
{\rm{Re}}z (\Theta_0^{\ 1}\Theta_1^{\ 1}+\Theta^{0 1}\Theta^{1 1}+\Theta_0^{\ 2}\Theta_1^{\ 2}+\Theta^{0 2}\Theta^{1 2}+\Theta_0^{\ 3}\Theta_1^{\ 3}+\Theta^{0 3}\Theta^{1 3})=0. \label{V=0_z}
\end{align}
Since $g^{z\bar{z}}=4({\rm{Re}}z)^2$, we should impose ${\rm{Re}}z\neq 0$, and the equation~$\eqref{V=0_z}$ constrains the components of the embedding tensor.\footnote{Under the condition~$\eqref{V=0_z}$, the scalar potential is exactly zero, and $z$ is a modulus.}

Next, let us consider some examples by assuming the following forms of the embedding tensors,\footnote{As we saw before, just one isometry gauging always leads to the degenerate breaking scale, thus we consider gauging two directions $(m=1,2)$ characterized by two real parameters.}
\begin{align}
\nonumber &(i)\ \ \Theta_{M}^{\ m}=\left(\begin{array}{ccc} E_1& E_2&0 \\ 0& 0& 0\\0& 0& 0\\0& 0& 0\\ \end{array} \right),\ \ (ii)\ \ \Theta_{M}^{\ m}=\left(\begin{array}{ccc} E_1& 0&0 \\ 0& E_2& 0\\0& 0& 0\\0& 0& 0\\ \end{array} \right),\\
&(iii)\ \ \Theta_{M}^{\ m}=\left(\begin{array}{ccc} E& 0&0 \\ 0& 0& 0\\0& 0& 0\\0& M& 0\\ \end{array} \right),
\end{align}
with all elements being real. All of the examples manifestly satisfy Eq.~$\eqref{V=0_z}$ as well as Eqs.~$\eqref{ETconst12}$-$\eqref{ETconst31}$. Then, we obtain the following expressions of $\sigma_A$, 
\begin{align}
(i)\ \ &\sigma_1=\sigma_2=\sqrt{E_1^2+E_2^2},\\
\nonumber (ii)\ \ &\sigma_1=\sqrt{E_1^2+E_2^2|z|^2-2E_1E_2{\rm{Im}}z}, \\
&\sigma_2=\sqrt{E_1^2+E_2^2|z|^2+2E_1E_2{\rm{Im}}z},\label{(ii)}\\
(iii)\ \ &\sigma_1=E-M,\ \ \sigma_2=E+M,
\end{align}
where we have assumed $E\geq M \geq 0$ in the example $(iii)$. The first example $(i)$ obviously predicts a degenerate breaking scale. For the examples $(ii)$ and $(iii)$, we show the gauge coupling dependence of the two breaking scales in Fig.~\ref{fig(ii,iii)}, changing the parameters as $0\leq E_1,E_2,E,M\leq 0.5$.
\begin{figure}[t]
 \begin{minipage}{0.5\hsize}
  \begin{center}
   \includegraphics[width=60mm]{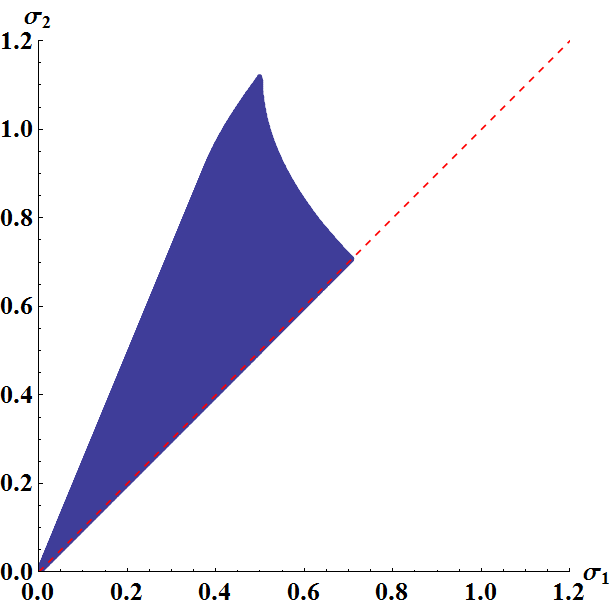}
  \end{center}
 \end{minipage}
 \begin{minipage}{0.5\hsize}
  \begin{center}
   \includegraphics[width=60mm]{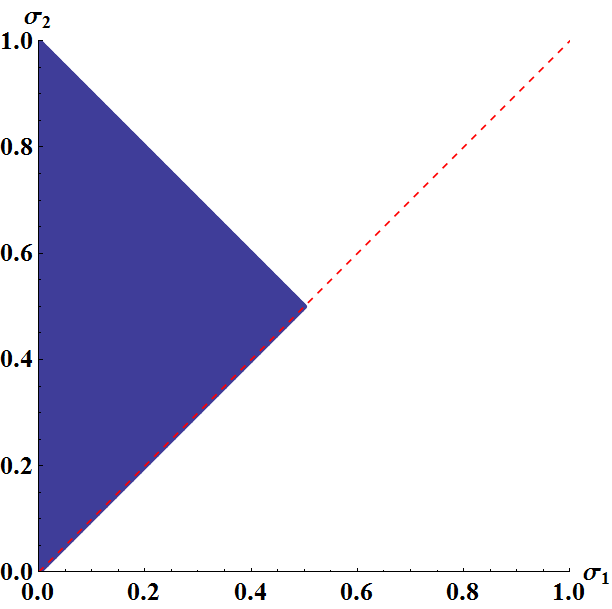}
  \end{center}
 \end{minipage}
  \caption{{\it{Left}} : The gauge coupling dependence of $\sigma_1$ and $\sigma_2$ in the case $(ii)$. We change $E_1,E_2$ as $0\leq E_1,E_2 \leq 0.5$ and set $z=1+i$. {\it{Right}} : The gauge coupling dependence of $\sigma_1$ and $\sigma_2$ in the case $(iii)$. We change $E,M$ as $0\leq E,M \leq 0.5$. The dotted line denotes $\sigma_2=\sigma_1$ in both figures.}
  \label{fig(ii,iii)}
\end{figure}
 In the case $(ii)$, we fixed $z=1+i$. In both cases, the two breaking scales can take different values, but the case $(ii)$ cannot cover the line $\sigma_1=0$ except for the origin, in contrast to the case $(iii)$. Note that this fact is independent of the value of $z$ in Eq.~$\eqref{(ii)}$ because $\sigma_1=0$ implies
\begin{align}
\nonumber \sigma_1=0\ \  &\Longleftrightarrow \ \ (E_1-E_2{\rm{Im}}z)^2+(E_2{\rm{Re}}z)^2=0 \\
&\Longleftrightarrow \ \ E_1=E_2=0,
\end{align}
and therefore, it also leads to $\sigma_2=0$ ($\mathcal{N}=2$ preserving vacuum). In the second equivalence, we have used ${\rm{Re}}z\neq 0$. This is consistent with the result of subsection.~\ref{N=1}, where it is explicitly shown that the pure electric gauging cannot realize the partial breaking $\sigma_1=0, \sigma_2\neq 0$. 

\begin{figure}[t]
 \begin{minipage}{0.5\hsize}
  \begin{center}
   \includegraphics[width=60mm]{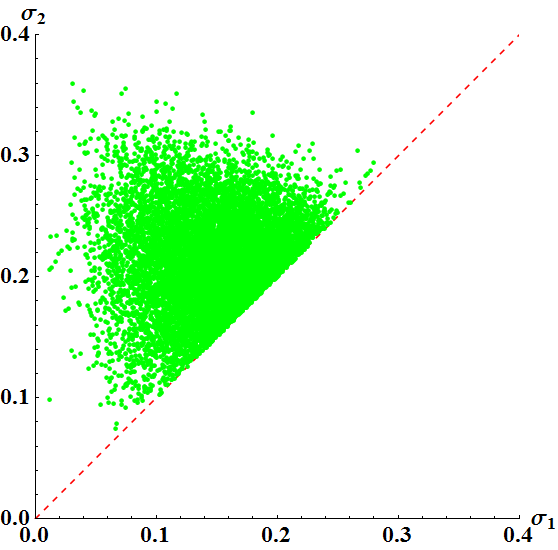}
  \end{center}
 \end{minipage}
 \begin{minipage}{0.5\hsize}
  \begin{center}
   \includegraphics[width=60mm]{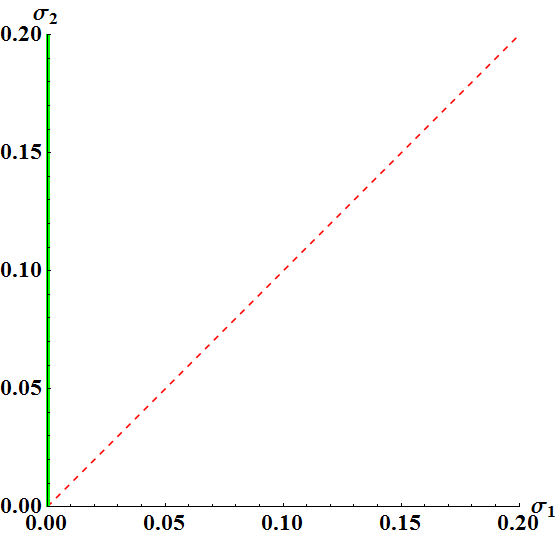}
  \end{center}
 \end{minipage}
 \caption{{\it{Left}} : The scatter plot of $\sigma_1$ and $\sigma_2$ in the case A. The value of $z$ is set to be $1$.  {\it{Right}} : The scatter plot of $\sigma_1$ and $\sigma_2$ in the case B. We have set $b=c=1$. In both figures, the gauge couplings in the embedding tensor are assigned to take the values in $\{-0.1,0.1\}$ and there are $10^4$ sample points.}
 \label{scatter}
\end{figure}

More generally, we plotted the values of $\sigma_A$ in Fig.~\ref{scatter} (the left) by randomly choosing the components of $\Theta_M^{\ m}$ in such a way that they satisfy the condition $\eqref{V=0_z}$ and the constraints $\eqref{ETconst12}$-$\eqref{ETconst31}$. 
All the components of $\Theta_M^{\ m}$ are assumed to take the values between $\{-0.1,0.1\}$ and $z$ is set to be $1$.


\subsubsection{Case B}
When $f_{zzz}\neq 0$ hold at the vacuum, the condition~$\eqref{V_i=02}$ must be satisfied. This condition is nothing but the partial breaking condition~$\eqref{PBcond3}$ with $n_v=1$. Therefore, $\mathcal{N}=1$ supersymmetry always remains at the vacuum in this case.\footnote{In Ref.~\cite{Antoniadis:2018blk}, this claim is proved by taking a concrete choice of the embedding tensor (see section~4.3 and Appendix~D of the paper). 
} Here, we explicitly construct a model satisfying the condition~$\eqref{V_i=02}$ and show that the partial breaking actually occurs.

Let us assume that the prepotential takes the form,
\begin{align}
f=az+bz^2+cz^3, \label{f_cub}
\end{align}
where $a,b$ and $c\neq 0$ are complex in general. For the condition~$\eqref{V_i=02}$ to have a nontrivial solution, we need to gauge at least two isometries and introduce the magnetic component as shown in subsection.~\ref{N=1}. For example, let us assume the following form of the embedding tensor, 
\begin{align}
\Theta_{M}^{\ m}=\left(\begin{array}{ccc} E_1& E_3&0 \\ E_2& 0& 0\\0& 0& 0\\ M& 0& 0\\ \end{array} \right),\label{assume_theta}
\end{align}
which satisfies Eqs.~$\eqref{ETconst12}$-$\eqref{ETconst31}$. Under this choice, one can check that the stationary conditions~$\eqref{V=0_single}$ and~$\eqref{V_i=02}$ are satisfied if
\begin{align}
&E_2-if_{zz}M=0,\\
&E_1+zE_2-iMf_z-iE_3=0,
\end{align}
are satisfied. These equations determine $z$ and $a$ as  
\begin{align}
z=-\frac{b}{3c}-i\frac{E_2}{6cM}, \ \ a=\frac{b^2}{3c}-\frac{E_3}{M}-\frac{E_2^2}{12cM^2}+i\left(-\frac{E_1}{M}+\frac{bE_2}{3cM}\right),\label{za_vev}
\end{align}
where we have assumed $b$ and $c$ are real just for simplicity.

In Fig.~\ref{scatter} (the right), we plotted the two breaking scales $\sigma_A$ under the conditions~$\eqref{za_vev}$.
The parameters $E_1,E_2,E_3$ and $M$ are assigned to take the values in $\{-0.1,0.1\}$, Also, we have set $b=c=1$.
It can be found that all the points are located on the line $\sigma_1=0$, which means that the partial breaking always occurs in this case. As explicitly shown in Ref.~\cite{Louis:2009xd}, the partially broken vacuum is ensured to be stable.


\subsection{Multiple vector multiplets}
Finally, we study the case of multiple vector multiplets. The case analysis of the condition~$\eqref{Vi=0}$ is not simple unlike the single case. Nevertheless, we roughly divide the situations into the following two cases,
\begin{align}
&{\rm{Case\ A:}}\ \ f_{ijk}=0,\ \ {\rm{for\ all\ }}i,\\
&{\rm{Case\ B}:}\ \ f_{ijk}\neq 0,\ \ {\rm{for\ some\ }}i,
\end{align}
in order to illustrate the similarity and the difference with $n_v=1$ case. 

The case A manifestly satisfies the condition~$\eqref{Vi=0}$. Then, all we have to take into account are only the constraints~$\eqref{ETconst12}$-$\eqref{ETconst31}$ and the Minkowski condition~$\eqref{Mcond}$. The situation is almost the same with the single case, and we can realize several types of supersymmetry breaking, $\mathcal{N}=0,1,2$.

In the case B, however, more conditions on the embedding tensor are required. The condition~$\eqref{Vi=0}$ in this case seems complicated, but we realize soon that it can be satisfied if the partial breaking condition~$\eqref{PBcond3}$ is satisfied, since 
\begin{align}
\nonumber &\nabla_j \alpha \nabla_k \alpha +\nabla_j \beta \nabla_k \beta +\nabla_j \gamma \nabla_k \gamma\\ &=\biggl[\frac{1}{4}\left(w-\frac{1}{w}\right)^2-\frac{1}{4}\left(w+\frac{1}{w}\right)^2 +1\biggr] \nabla_j \gamma \nabla_k \gamma=0.
\end{align}
Therefore, we can obtain the partially broken vacuum as $n_v=1$ case.

However, there exist other solutions of Eq.~$\eqref{Vi=0}$ which do not necessarily satisfy the partial breaking condition. This is contrast to the situation of $n_v=1$ case, where we always have the partially broken vacuum if the cubic coupling in the prepotential exists. To show it based on an explicit model, we consider $n_v=2$ case, and choose the prepotential as 
\begin{align}
f(z_1,z_2)=z_1^2z_2.
\end{align}
Also, the form of the embedding tensor is assumed to be 
\begin{align}
\Theta_{M}^{\ m}=\left(\begin{array}{ccc} \Theta_{0}^{\ 1}& \Theta_{0}^{\ 2}&0 \\ \Theta_{1}^{\ 1}& 0& 0\\0& \Theta_{2}^{\ 2}& 0\\ 0& 0& 0\\ \Theta^{11}& 0& 0\\ 0& \Theta^{22}& 0\\ \end{array} \right),
\end{align}
which satisfies the constraints~$\eqref{ETconst12}$-$\eqref{ETconst31}$.
Then, we found that the sets 
\begin{align}
&{\rm{(I)}}\ \ z_1=z_2=1+i, \ \ \Theta_{0}^{\ 2}=2\Theta^{22}, \ \ \Theta_{0}^{\ 1}=\Theta_{1}^{\ 1}=\Theta_{2}^{\ 2}=0,\\
&{\rm{(II)}}\ \ z_1=z_2=1+i, \ \ \Theta_{0}^{\ 2}=2\Theta^{22}, \ \ \Theta_{0}^{\ 1}=\Theta_{2}^{\ 2}=\Theta^{11}=0,\\
&{\rm{(III)}}\ \ z_1=z_2=1+i, \ \ \Theta_{0}^{\ 1}=\Theta_{0}^{\ 2}=\Theta_{1}^{\ 1}=\Theta_{2}^{\ 2}=\Theta^{22}=0,\\
&{\rm{(IV)}}\ \ z_1=z_2=1+i, \ \ \Theta_{0}^{\ 1}=\Theta_{0}^{\ 2}=\Theta_{2}^{\ 2}=\Theta^{11}=\Theta^{22}=0,,
\end{align}
are the solutions of Eqs.~$\eqref{Mcond}$ and~$\eqref{Vi=0}$. Each value of $\sigma_A$ is given by 
\begin{align}
&{\rm{(I)}}\ \ \sigma_1=\sigma_2=X_{{\rm{(I)}}+},\\
&{\rm{(II)}}\ \ \sigma_1=X_{{\rm{(II)}}+}-X_{{\rm{(II)}}-},\ \ \sigma_2=X_{{\rm{(II)}}+}+X_{{\rm{(II)}}-},\label{sigma_II}\\
&{\rm{(III)}}\ \ \sigma_1=\sigma_2=4|\Theta^{11}|,\\
&{\rm{(IV)}}\ \ \sigma_1=\sigma_2=4|\Theta_1^{\ 1}|,
\end{align}
with
\begin{align}
&X_{{\rm{(I)}}+}=4\sqrt{(\Theta^{11})^2+(\Theta^{22})^2},\ \ X_{{\rm{(I)}}-}=0,\\
&X_{{\rm{(II)}}+}=\sqrt{(\Theta_{1}^{\ 1})^2+8(\Theta^{22})^2+\sqrt{(\Theta_{1}^{\ 1})^4+64(\Theta^{22})^4}},\label{X_+_II}\\
&X_{{\rm{(II)}}-}=\sqrt{(\Theta_{1}^{\ 1})^2+8(\Theta^{22})^2-\sqrt{(\Theta_{1}^{\ 1})^4+64(\Theta^{22})^4}}.\label{X_-_II}
\end{align}
Here $X_{\pm}$ are defined by Eq.~$\eqref{Xpm}$. Then, it can be found that $\sigma_1$ can be nonzero, and there exist full broken vacua even when we consider the case $f_{ijk}\neq 0$. In the case (I),(III), and (IV), we have degenerate supersymmetry breaking scales. As for the case (II), we have shown the $\Theta_{1}^{\ 1}$ and $\Theta^{22}$ dependence in Fig.~\ref{listexample(n=2,case2,full)}.
\begin{figure}[t]
\hspace*{4cm}
\includegraphics[width=8.0cm]{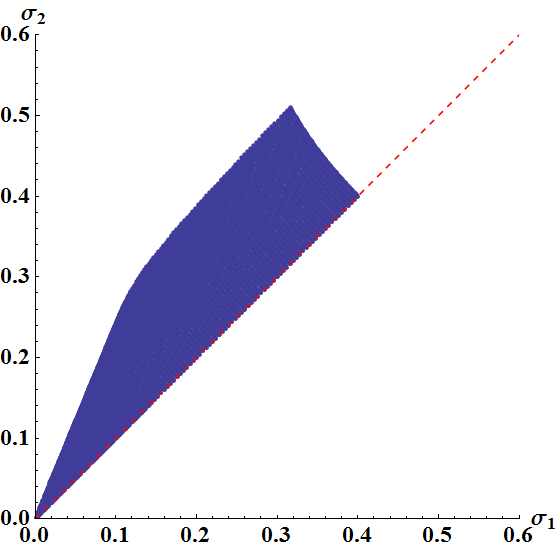}
\caption{The gauge coupling dependence of $\sigma_1$ and $\sigma_2$ in the case (II), which is given by Eqs.~$\eqref{sigma_II}$, $\eqref{X_+_II}$, and $\eqref{X_-_II}$. We change the gauge couplings, $\Theta_{1}^{\ 1}$ and $\Theta^{22}$, between $\{0,0.1\}$.}
\label{listexample(n=2,case2,full)}
\end{figure}
These parameters are assumed to take the values between $\{0,0.1\}$. 

As for the stability of the scalar potential, we evaluate eigenvalues of the Hessian matrices and obtain
\begin{align}
&{\rm{(I)}}\ \ \biggl\{\frac{1}{2}\left(\frac{X_{{\rm{(I)}}+}}{4}\right)^2-\frac{1}{2}\sqrt{(\Theta^{11})^4+(\Theta^{22})^4},\frac{1}{2}\left(\frac{X_{{\rm{(I)}}+}}{4}\right)^2+\frac{1}{2}\sqrt{(\Theta^{11})^4+(\Theta^{22})^4},0,0\biggr\},\\
&{\rm{(II)}}\ \ \biggl\{\frac{1}{4}X_{{\rm{(II)}}-}^2,\frac{1}{4}X_{{\rm{(II)}}+}^2,0,0 \biggr\},\\
&{\rm{(III)}}\ \ \biggl\{4(\Theta^{11})^2,0,0,0 \biggr\},\\
&{\rm{(IV)}}\ \ \biggl\{\frac{1}{2}(\Theta_1^{\ 1})^2,0,0,0 \biggr\}.
\end{align}
Although there are some massless scalars, tachyonic mode does not exist.

\section{Summary}\label{Summary}
In the paper, we analyzed the patterns of supersymmetry breaking in $\mathcal{N}=2$ gauged supergravity with multiple vector multiplets and a single hypermultiplet. Based on the embedding tensor formalism, we derived the general expressions of the two gravitino masses~$\eqref{sigma12}$ (supersymmetry breaking scales) under the gauging of the isometry~$\eqref{shift}$. Then, we discussed how they change depending on the input parameters such as the gauge coupling constants and the prepotential, taking into account the conditions the vacuum must satisfy.

In the case with a single vector multiplet, we can classify the situation by the vacuum expectation value of the third derivative of the prepotential, $f_{zzz}$. When $f_{zzz}= 0$, we have varieties of the breaking patterns, depending on the gauge couplings (see Fig.~\ref{fig(ii,iii)}). When $f_{zzz}\neq 0$, on the other hand, it was shown that the $\mathcal{N}=1 $ supersymmetry always remains, which is consistent with the previous result of Ref.~\cite{Antoniadis:2018blk}. This result does not depend on the specific choice of the prepotential and the form of the embedding tensor, as long as Eq.~$\eqref{V_i=02}$ has a solution. 
For the case of multiple vector multiplets, we found that the full breaking can be realized even when the third derivatives of the prepotential are nontrivial. These observations would be important when we discuss the relation to the string compactifications, D-brane effective action, and the particle phenomenology/cosmology.  

As future directions, it is important to investigate the mass spectrum other than the gravitinos, especially, how they change depending on the two supersymmetry breaking scales and affects the low energy physics. There is also a room for further generalizations of our model: The extension of the hyper sector and non-Abelian generalization may change the situation significantly. Also, applications to other extended supergravities in various dimensions are interesting themes. We will study these issues elsewhere.


\subsection*{Acknowledgements}
S.~A. would like to thank Henry Liao for useful discussion. H.~A. and S.~A. are supported in part by a Waseda University Grant for Special Research Projects (Project
number: 2019Q-027 and 2019E-059, respectively). H.~A. is also supported by Institute for Advanced Theoretical and Experimental Physics, Waseda University.
\begin{appendix}
\section{Spinor notation} \label{notation}

Here, we summarize spinor conventions.

The SU(2) and Sp(2) invariant tensors satisfy
\begin{align}
&\epsilon^{AB}\epsilon_{BC}=-\delta^A_C, \ \ \epsilon^{12}=\epsilon_{12}=1,\\
&\mathbb{C}^{\alpha \beta}\mathbb{C}_{\beta \gamma}=-\delta^{\alpha}_{\gamma},\ \ \mathbb{C}^{12}=\mathbb{C}_{12}=1,
\end{align}
and the indices of SU(2) and Sp(2) vectors are raised and lowered by
\begin{align}
\epsilon_{AB}P^B=P_A,\ \ \epsilon^{AB}P_B=-P^A,\\
\mathbb{C}_{\alpha \beta}P^{\beta}=P_{\alpha},\ \ \mathbb{C}^{\alpha \beta}P_{\beta}=-P^{\alpha}.
\end{align}

The Pauli matrices are $(\tau^x)_A^{\ B}(x=1,2,3)$ are
\begin{align}
(\tau^1)_A^{\ B}=\left(\begin{array}{cc}  0&1 \\   1& 0\\ \end{array} \right), \ \ (\tau^2)_A^{\ B}=\left(\begin{array}{cc}  0&-i \\   i& 0\\ \end{array} \right), \ \ (\tau^3)_A^{\ B}=\left(\begin{array}{cc}  1&0 \\   0& -1\\ \end{array} \right).   
\end{align}
Their indices are raised and lowered by $\epsilon_{AB}$ and $\epsilon^{AB}$ defined above.

We denote the chirality of the spinors as 
\begin{align}
&\gamma_5\left( \begin{array}{cc} \psi_A\\ \lambda^{iA}\\ \zeta_{\alpha}\\ \epsilon_A\\ \end{array} \right)=\left( \begin{array}{cc} \psi_A\\ \lambda^{iA}\\ \zeta_{\alpha}\\ \epsilon_A\\ \end{array} \right), \\
&\gamma_5\left( \begin{array}{cc} \psi^A\\ \lambda^{\bar{i}}_{A}\\ \zeta^{\alpha}\\ \epsilon^A\\ \end{array} \right)=-\left( \begin{array}{cc} \psi^A\\ \lambda^{\bar{i}}_{A}\\ \zeta^{\alpha}\\ \epsilon^A\\ \end{array} \right).
\end{align}



\end{appendix}


\end{document}